\documentclass[letterpaper]{article}
\usepackage{aaai}
\usepackage{times}
\usepackage{helvet}
\usepackage{courier,url}
\usepackage{graphicx,pifont,multirow,diagbox}
\usepackage{color,booktabs}
\usepackage{float,makecell}
\usepackage{verbatim}

\usepackage[most]{tcolorbox}
\usepackage{lipsum}

\usepackage{subfigure}

\usepackage{enumitem}
\usepackage{array}
\newcolumntype{H}{>{\setbox0=\hbox\bgroup}c<{\egroup}@{}}

\newcommand{\suhang}[1]{\textcolor{blue}{Suhang: {#1}}}

\newcommand{\TODO}[1]{\textcolor{cyan}{Deepak TODO: {#1}}}

\begin{document}
%
\title{Hierarchical Propagation Networks for Fake News Detection: \\Investigation and Exploitation}
\author{Kai Shu$^{1}$, Deepak Mahudeswaran$^{1}$, Suhang Wang$^{2}$, and  Huan Liu$^{1}$ \vspace{0.1in} \\
$^{1}$Computer Science and Engineering, Arizona State University, Tempe, 85281, USA\\
$^{2}$College of Information Sciences and Technology, Penn State University, University Park, PA, 16802, USA\\
\{kai.shu, dmahudes, huan.liu\}@asu.edu, szw494@psu.edu
}
\maketitle
\begin{abstract}

Consuming news from social media is becoming increasingly popular.  However, social media also enables the wide dissemination of \textit{fake news}. Because of the detrimental effects of fake news, fake news detection has attracted increasing attention. However, the performance of detecting fake news only from news content is generally limited as fake news pieces are written to mimic true news. In the real world, news pieces spread through \textit{propagation networks} on social media. The news propagation networks usually involve multi-levels.
In this paper, we study the challenging problem of \textit{investigating} and \textit{exploiting} news hierarchical propagation network on social media for fake news detection. 

In an attempt to understand the correlations between news propagation networks and fake news, first, we build a hierarchical propagation network from macro-level and micro-level of fake news and true news; second, we perform a comparison analysis of the propagation network features from structural, temporal, and linguistic perspectives between fake and real news, which demonstrates the potential of utilizing these features to detect fake news; third, we show the effectiveness of these propagation network features for fake news detection. We further validate the effectiveness of these features from feature important analysis. 
Altogether, this
work presents a data-driven view of hierarchical propagation network and fake news, and paves the way towards a healthier online news ecosystem.
\end{abstract}

\section{Introduction}
Social media platforms are easy to access, support fast dissemination of posts, and allow users to comment and share, which are attracting more and more users to seek out and receive timely news information online. For example, the Pew Research Center announced that approximately 68\% of US adults get news from social media in 2018, while in 2012, only 49\% reported seeing news on social media\footnote{http://www.journalism.org/2018/09/10/news-use-across-social-media-platforms-2018/}. 
However, social media also enables the wide dissemination of large amounts of fake news, i.e.,  news stories with intentionally false information~\cite{allcott2017social,shu2017fake}. For example, a report estimated that over 1 million tweets were related to the fake news story ``Pizzagate''~\footnote{https://en.wikipedia.org/wiki/Pizzagate\_conspiracy\_theory} by the end of 2016 presidential election.

The widespread of fake news has detrimental societal effects. First, it weakens the public trust in governments and journalism. For example, the reach of fake news during the 2016 U.S. presidential election campaign for top twenty fake news pieces was, ironically, larger than the top twenty most-discussed true stories~\footnote{https://www.buzzfeednews.com/article/craigsilverman/viral-fake-election-news-outperformed-real-news-on-facebook}.
Second, fake news may change the way people respond to legitimate news. A study has shown that people's trust to mass media has dropped dramatically across different age groups and political parties. ~\cite{swift2016americans}. Third, rampant fake news can lead to real-life societal events. For example, fake news claiming that Barack Obama was
injured in an explosion wiped out \$130 billion in stock value~\footnote{https://www.forbes.com/sites/kenrapoza/2017/02/26/can-fake-news-impact-the-stock-market/\#4986a6772fac}.
Thus, detecting fake news on social media is a precursor to mitigating negative effects, and promoting trust in the entire news ecosystem.

However, detecting fake news on social media presents unique challenges. First, fake news is intentionally written to mislead readers, which makes it nontrivial to detect simply based on content; Second, social media data is large-scale, multi-modal, mostly user-generated, sometimes anonymous and noisy. Recent research advancements aggregate uses' social engagements on news pieces to help infer which articles are fake~\cite{jin2016news,liu2018early}, giving some promising early results. For example, Jin \textit{et al.} propose to exploit users' conflicting viewpoints from social media posts and estimate their credibility values for fake news detection. Liu \textit{et al.} utilize a deep neural network model to classify the news propagation path constructed by tweets and retweets to detect fake news.

\begin{figure}[!tp]
\centering{
\hspace{-0.2cm}
\includegraphics[scale=0.25]{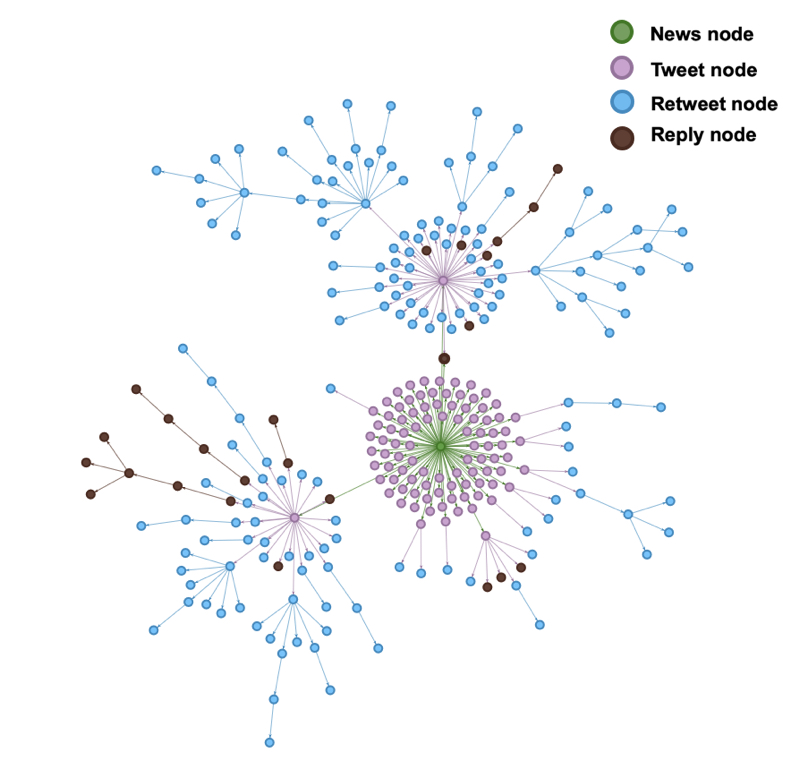}
\vskip -1em
\caption{An example of the hierarchical propagation network of a fake news piece fact-checked by Politifact~\footnotemark. It consists of two types: \textbf{macro-level} and \textbf{micro-level}. The macro-level propagation network includes the news nodes, tweet nodes, and retweet nodes. The micro-level propagation network indicates the conversation tree represented by reply nodes.}
\label{fig:crawler_workflow}}
\end{figure}

In the real world, news pieces spread in networks on social media. The propagation networks have a hierarchical structure, including macro-level and micro-level propagation networks (see Figure~\ref{fig:crawler_workflow}). On one hand, macro-level propagation networks demonstrate the spreading path from news to the social media posts sharing the news, and those reposts of these posts. Macro-level networks for fake news are shown to be deeper, wider, and includes more social bots than real news~\cite{shao2017spread,vosoughi2018spread}, which provides clues for detecting fake news. 
On the other hand, micro-level propagation networks illustrate the user conversations under the posts or reposts, such as replies/comments. Micro-level networks contain user discussions towards news pieces, which brings auxiliary cues such as sentiment polarities~\cite{gilbert2014vader}, stance signals~\cite{jin2016news}, to differentiate fake news.
Studying macro-level and micro-level propagation network provides fine-grained social signals to understand fake news and can possibly facilitate fake news detection. Despite the seminal work in analyzing the macro-level propagation network from temporal or structural perspectives~\cite{vosoughi2018spread}, no principled study is conducted on characterizing the propagation network from a hierarchical perspective on social media, let alone exploring whether/how these features can help fake news detection.
\footnotetext{https://bit.ly/2H8FnR5}
In addition, there is no research that actually provides a deep understanding of (i) how fake news and true news propagate differently from micro-level and macro-level; (ii) whether features extracted from hierarchical propagate networks are useful for fake news detection; and (iii) how discriminative these features are. To give a comprehensive understanding, we investigate the following two research questions:

\begin{itemize}
  \item \textbf{RQ1:} \emph{What are the characteristics of the structure, temporal and linguistic of hierarchical propagation networks of fake and real news?}
  \item \textbf{RQ2:} \emph{Can we use the extracted features to detect fake news and how?}
\end{itemize}
By investigating \textbf{RQ1}, we aim to assess whether the propagation network features of fake and real news are different or not from micro-level and macro-level, and to what extent and in what aspects they are different. In addition, by studying \textbf{RQ2}, we explore different ways to model propagation network features, analyze the importance of each feature, and show the feature robustness to various learning algorithms. By answering these research questions, we made the following contributions:
\begin{itemize}
  \item We study a novel problem of understanding the relationships between hierarchical propagation network and fake news, which lays the foundation of exploiting them for fake news detection;
  \item We propose a principled way to characterize and understand hierarchical  propagation network features. We perform a statistical comparative analysis over these features, including micro-level and macro-level, of fake news and true news; and 
  \item We demonstrate the usefulness of the extracted hierarchical network features to classify fake news, whose performance consistently outperforms the  existing state-of-the-art methods. We also show that the extracted propagation network features are robust to different learning algorithms, with an average $F1>0.80$.
 We further validate the effectiveness of these features through feature importance analysis, and found that temporal and structure features perform better than linguistic features.
  \end{itemize}

The remainder of this paper is organized as follows.  In Section~\ref{sec:construct}, we propose to construct the hierarchical propagation networks for news pieces. In Section~\ref{sec:charac}, we analyze the feature of macro-level and micro-level propagation networks. In Section~\ref{sec:eval}, we evaluate the effectiveness of the extracted propagation network features for fake news detection. In Section~\ref{sec:related}, we review related work. In Section~\ref{sec:conclude}, we conclude with future works.

\section{Constructing Propagation Networks}\label{sec:construct}
In this section, we investigate how to construct the hierarchical propagation networks of fake and real news.  We aim to explore how we can capture the news spreading process in a propagation network with different granularity such as micro-level and macro-level,  which can be further utilized to extract discriminative features from different perspectives for fake news detection. 

\subsection{Datasets}
\begin{table}[t]
\small
\centering \caption{The statistics of FakeNewsNet}
\begin{tabular}{l|cc}
\toprule
 Platform &PolitiFact & GossipCop\\
\midrule
\# True news &624 &16,817  \\
\midrule
\# Fake news &432   &5,323  \\
\midrule
 \makecell[l]{\# True news \\with propagation network}& 277  & 6,945\\
\midrule
\makecell[l]{\# Fake news \\with propagation network} & 351 & 3,684  \\
\midrule
\# Users & 384,813& 739,166\\
\midrule
\# Tweets & 275,058& 1,058,330 \\
\midrule
\# Retweets & 293,438& 530,833\\
\midrule
\# Replies & 125,654& 232,923\\

\bottomrule
\end{tabular} \label{tab:data}
\vskip -1em
\end{table}

We utilize the public fake news detection data repository FakeNewsNet~\cite{shu2017fake}. The repository consists of news data related to different fact-checking websites and the correspondent information of news content, social context, and dynamic information.

We use the data from following fact-checking websites: \textit{GossipCop} and \textit{PolitiFact}, both containing news content with labels annotated by professional journalists, social context, and temporal information. News content includes the meta attributes of the news (e.g., body text), the social context includes the related user social engagements of news items (e.g., user posting/sharing/commenting news in Twitter), and  dynamic information includes the timestamps of users' engagements. The detailed statistics of the datasets are shown in Table~\ref{tab:data}. Next, we introduce how to build hierarchical  propagation networks for fake and real news from FakeNewsNet.

\subsection{Hierarchical Propagation Networks}

The hierarchical propagation network is constructed in different levels of granularity including \textbf{micro-level} and \textbf{macro-level}. Micro-level networks represent the network of replies where information is shared in the local level. Macro-level networks represent global propagation of information in Twitter through a cascade of retweets. Through hierarchical propagation network, both local and global pattern of information diffusion related to fake and real news can be studied.

For macro-level propagation network, nodes represent the tweets and the edges represent the retweet relationship among them. In a macro network, an edge exists from node $u$ to $v$ when a tweet $u$ is retweeted by some user $x$ and node $v$ is created as a result of it. In Twitter, a tweet or a retweet can be retweeted. However, in the retweet data collected from official Twitter API, there is no indication whether retweeted sources is an original tweet or another retweet. So the retweet network cannot be explicitly constructed from the data available from official Twitter API data. Hence a different strategy using social network \cite{goel2015structural} of the users is used to construct a macro propagation network. For inferring the source of the retweet, we can identify the potential user's friends who retweeted the tweet. If the timestamp of the user's retweets is greater than the time stamp of the one of the user friend's retweet time stamp, then the user must have mostly seen the tweet from one of his/her friends and retweeted it. In a case where immediate retweet from a user's friend is not found, we can consider the retweet is done from the original tweet rather than retweet of another retweet.

For the micro-level propagation network, the nodes representing the replies to the tweets posting news and edges represent the relationship among them. In Twitter, a user can reply to actual tweet or reply of another user. In cases where user replies to the original tweet, then an edge is between tweet posting news and the current node. In case where users reply to the reply of another user, a conversation thread is formed and this is represented as the chain of replies in the propagation path.




\section{Characterizing Propagation Networks}\label{sec:charac}
In this section, we address \textbf{RQ1} by performing a comparison analysis on the constructed hierarchical propagation networks for fake news and real news from different perspectives.

\begin{table*}[!tbp]
\small
\label{tab:macro_struct}
\begin{center}
\caption{Statistics of structural features for macro propagation network. Stars denotes statistically significant under $t$-test.}
\label{tab:baseline_perf}
\begin{tabular}{| c | c | c | c| c| c|c | c |c | c|c|c|c|}
\hline
\textbf{Features } & \multicolumn{6}{ c |} {\textbf{PolitiFact}} & \multicolumn{6}{  c |}{\textbf{GossipCop}}\\ 
\cline{2-13}
& \multicolumn{3}{ c |}{\textbf{Fake}} & \multicolumn{3}{  c |}{\textbf{Real}} &  \multicolumn{3}{ c |}{\textbf{Fake}} & \multicolumn{3}{  c |}{\textbf{Real}}\\
\cline{2-13}
& \textbf{Min} & \textbf{Max} & \textbf{Avg}& \textbf{Min} & \textbf{Max} & \textbf{Avg} & \textbf{Min} & \textbf{Max} & \textbf{Avg}& \textbf{Min} & \textbf{Max} & \textbf{Avg}\\\hline
$\mathbf{S_{1}}$ &2 & 14 &  5.93 $^{*}$ & 2 & 13 & 5.49$^{*}$ & 2 &12 &3.89 $^{*}$ &2& 10 & 3.43$^{*}$ \\
\hline
$\mathbf{S_{2}}$& 2 & 35,189 & 774.65$^{*}$ & 2 & 23,494 & 1,205.46$^{*}$ &2 & 5339 & 272.14 $^{*}$  & 2 & 2,497 & 108.76 $^{*}$\\
\hline
$S_{3}$& 0 & 145 & 27.37 & 0 & 95 & 31.35  &0&  198 & 14.42 $^{*}$ & 0 & 98 & 12.44 $^{*}$\\ 
\hline
$S_{4}$ & 1 & 17,548 & 415.59 & 1 & 9,577 &  537.0 & 1 & 2568 &158.67 $^{*}$ & 1 & 1625 & 80.19 $^{*}$  \\
\hline
$\mathbf{S_{5}}$& 0 & 7 & 1.17$^{*}$ & 0 & 5 & 1.03$^{*}$ &0 & 4 & 1.15 $^{*}$ & 0 &6  &0.94 $^{*}$ \\
\hline
$S_{6}$& 0 & 3207 & 56.79 & 0 & 1640 & 84.55 & 0 & 421 & 18.57$^{*}$ & 0 & 214& 3.58$^{*}$\\
\hline
$\mathbf{S_{7}}$& 0 & 1 & 0.16$^{*}$ &  0 & 1 & 0.08$^{*}$ & 0 & 1 & 0.15$^{*}$ & 0 & 1 & 0.05$^{*}$ \\
\hline
$S_{8}$& 0 & 2462 & 63.26 & 0 &  1735 & 77.05 & 0 & 461 & 12.86$^{*}$ & 0 & 125 & 3.37$^{*}$ \\
\hline
$S_{9}$&0.01  & 0.68& 0.23 &0.03 &0.8 & 0.21& 0.01 &  0.89 &  0.29 $^{*}$& 0.01 & 0.97 & 0.32$^{*}$ \\
\hline
\end{tabular}
\end{center}
\vskip -1em
\end{table*}

\subsection{Macro-Level Propagation Network}

Macro-level propagation network encompasses information on tweets posting pattern and information sharing pattern. We analyze the macro-level propagation network in terms of structure and temporal aspects. Since the same textual information related to a news article is shared across the macro-level network, linguistic analysis is not applicable.

\subsubsection{Structural analysis}
Structural analysis of macro-level networks helps to understand the global spreading pattern of the news pieces.  Existing work has shown that learning latent features from the macro-level propagation paths can help to improve fake news detection, while lacking of an in-depth understanding of why and how it is helpful~\cite{wu2018tracing,liu2018early}. Thus, we characterize and compare the macro-level propagation networks by looking at various network features as follows.



\begin{itemize}

\item  $(\mathbf{S_{1}})$ \textit{Tree depth}: 
The depth of the macro propagation network, capturing how far the information is spread/retweeted by users in social media. 


\item $(\mathbf{S_{2}})$ \textit{Number of nodes}: The number of nodes in a macro network indicates the number of users who share the new article and can be a signal for understanding the spreading pattern. 


\item  $(\mathbf{S_{3}})$ \textit{Maximum Outdegree}: Maximum outdegree in macro network could reveal the tweet/retweet with the most influence in the propagation process. 

\item $(\mathbf{S_{4}})$ \textit{Number of cascades}: The number of original tweets posting the original news article.

\item $(\mathbf{S_{5}})$ \textit{Depth of node with maximum outdegree}: The depth at which node with maximum outdegree occurs. This indicates steps of propagation it takes for a news piece to be spread by an influential node whose post is retweeted by more users than any other user's repost.

\item $(\mathbf{S_{6}})$ \textit{Number of cascades with retweets}: It indicate number of cascades (tweets) those were retweeted at least once. 

\item $(\mathbf{S_{7}})$ \textit{Fraction of cascades with retweets}: It indicates the fraction of tweets with retweets among all the cascades.

\item $(\mathbf{S_{8}})$ \textit{Number of bot users retweeting}: This feature captures the number of bot users who retweet the corresponding news pieces.

\item $(\mathbf{S_{9}})$ \textit{Fraction of bot users retweeting}: It is the ratio of bot users among all the users who tweeting and retweeting a news piece. This feature can show whether news pieces are more likely to be disseminated by bots or real humans.



\end{itemize}

We obtain the aforementioned structural features for macro-level propagation networks of fake news and real news in both Politifact and Gossipcop datasets. As shown in Table~\ref{tab:macro_struct}, we analyze the distribution of structural features and have the following observations: 
\begin{itemize}
    \item The features $S_1, S_2$, $S_5$ and $S_7$ are consistently different from fake news and real news in both datasets, under the statistical $t$-test.
    \item The average depth of the macro-level propagation network ($S_1$) of fake news is larger than that of real news in both PolitiFact and GossipCop significantly. This shows fake news has a longer chain of retweets than real news, which is consistent with the observation in~\cite{vosoughi2018spread}.
    \item Further, the depth of the node with the maximum out-degree ($S_5$) of fake news is greater than that of real news on both datasets, which indicates fake news takes longer steps to be reposted by an influential user.
    \item We can see that the fraction of cascades with retweets is larger for macro-level propagation network for fake news than that for real news. It shows that there are more number of tweets posting fake news are retweeted in average.
\end{itemize}

\subsubsection{Temporal analysis} 
\begin{table*}[!htbp]
\small
\label{tab:micro_struct}
\begin{center}
\caption{Statistics of structural features for micro propagation network. Stars denote statistically significant under $t$-test.}
\label{tab:baseline_perf}
\begin{tabular}{| c | c | c | c| c| c| c | c |c | c|c|c|c|}
\hline
\textbf{Features } & \multicolumn{6}{ c |} {\textbf{PolitiFact}} & \multicolumn{6}{  c |}{\textbf{GossipCop}}\\ 
\cline{2-13}
& \multicolumn{3}{ c |}{\textbf{Fake}} & \multicolumn{3}{  c |}{\textbf{Real}} &  \multicolumn{3}{ c |}{\textbf{Fake}} & \multicolumn{3}{  c |}{\textbf{Real}}\\
\cline{2-13}
& \textbf{Min} & \textbf{Max} & \textbf{Avg}& \textbf{Min} & \textbf{Max} & \textbf{Avg} & \textbf{Min} & \textbf{Max} & \textbf{Avg}& \textbf{Min} & \textbf{Max} & \textbf{Avg}\\\hline
$\mathbf{S_{10}}$ & 2 & 6 & 4.57$^{*}$ & 2 & 6 &  4.25$^{*}$ & 2 & 6 & 3.40$^{*}$ & 2& 6 &2.51$^{*}$ \\
\hline
$\mathbf{S_{11}}$& 2 & 21,923 & 544.91$^{*}$ & 2 & 18,522 & 853.89$^{*}$ & 2 & 3,453 & 213.97$^{*}$ &2 &  1696 & 90.75$^{*}$\\
\hline
$S_{12}$& 0& 204 & 24.29& 0 & 210 & 28.45 & 0 & 234 & 9.82 &0 &191 & 4.54 \\ 
\hline
$S_{13}$&0  & 1089 & 26.29$^{*}$& 0 &1185 & 45.33 $^{*}$ &   0 & 401 &12.36$^{*}$ &0 & 145& 1.64$^{*}$ \\ 
\hline
$\mathbf{S_{14}}$&  0 & 1  & 0.09$^{*}$&0  &  1& 0.06$^{*}$ &  0 &1  & 0.06$^{*}$ & 0 & 1 & 0.02$^{*}$  \\ 
\hline
\end{tabular}
\end{center}
\vskip -1em
\end{table*}


The temporal user engagements in macro-level network reveal the frequency and intensity of news dissemination process. The frequency distribution of user posting over time can be encoded in recurrent neural networks to learn the features to detection fake news~\cite{ruchansky2017csi,shu2018fakenewstracker}. However, the learned features are not interpretable, and the explanation of why the learned features can help remain unclear. Here, we extract several temporal features from macro-level propagation networks explicitly for more explainable abilities and analyze whether these features are different or not. Following are the features we extracted from the macro propagation network,
\begin{itemize}

\item $(\mathbf{T_{1}})$ \textit{Average time difference between the adjacent retweet nodes}: It indicates how fast the tweets are retweeted in news dissemination process.


\item $(\mathbf{T_{2}})$ \textit{Time difference between the first tweet and the last retweets}: It captures the life span of the news spread process.



\item $(\mathbf{T_{3}})$ \textit{Time difference between the first tweet and the tweet with maximum outdegree}: Tweets with maximum outdegree in propagation network represent the most influential node. This feature demonstrates how long it took for a news article to be retweeted by most influential node.


\item $(\mathbf{T_{4}})$ \textit{Time difference between the first and last tweet posting news}: This indicates how long the tweets related to a news article are posted in Twitter.

\item $(\mathbf{T_{5}})$ \textit{Time difference between the tweet posting news and last retweet node in deepest cascade}: Deepest cascade represents the most propagated network in the entire propagation network. This time difference indicates the lifespan of the news in the deepest cascade and can show whether news grows in a bursty or slow manner.

\item $(\mathbf{T_{6}})$ \textit{Average time difference between the adjacent retweet nodes in the deepest cascade}: This feature indicates how frequent a news article is retweeted in the deepest cascade.

\item $(\mathbf{T_{7}})$ \textit{Average time between tweets posting news}: This time indicates whether tweets are posted in short interval related to a news article.

\item $(\mathbf{T_{8}})$ \textit{Average time difference between the tweet post time and the first retweet time}: The average time difference between the first tweets and the first retweet node in each cascade can indicate how soon the tweets are retweeted. 




\end{itemize}

\begin{figure*}[!t]
\label{fig:temp_politifact_boxplot}
\begin{tcbraster}[raster columns=6, enhanced, blankest]
\tcbincludegraphics[scale=1.05]{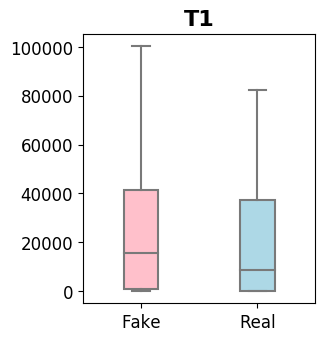}
\tcbincludegraphics{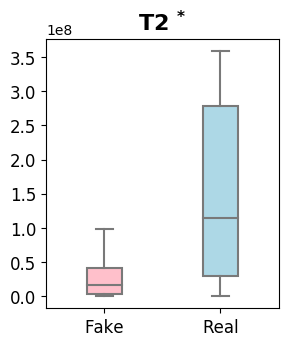}
\tcbincludegraphics{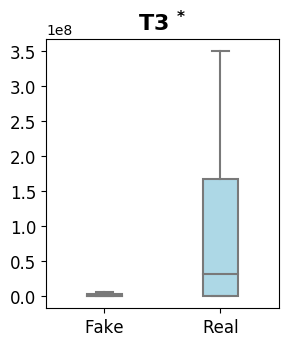}
\tcbincludegraphics{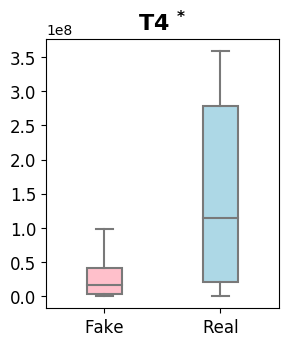}
\tcbincludegraphics[scale=1.04]{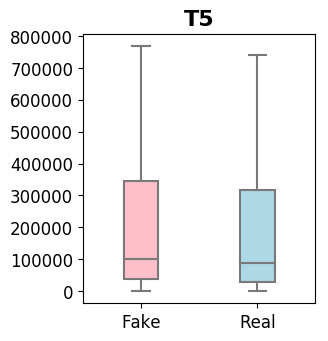}
\tcbincludegraphics[scale=1.035]{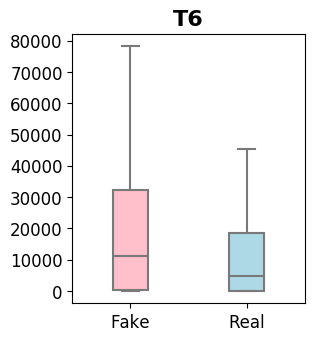}
\tcbincludegraphics{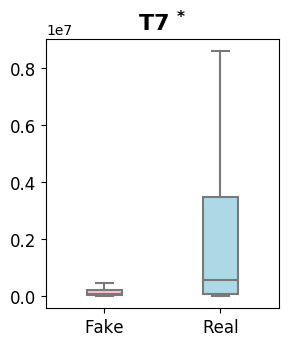}
\tcbincludegraphics{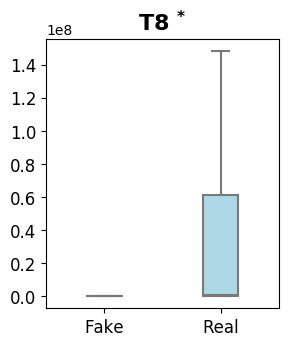}
\tcbincludegraphics[scale=1.05]{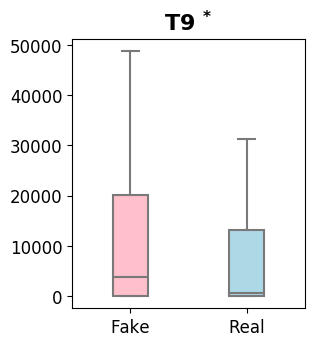}
\tcbincludegraphics{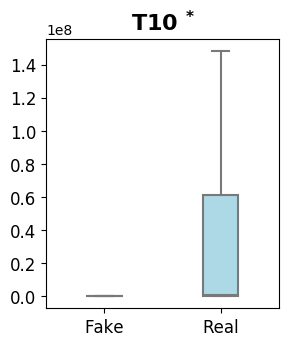}
\tcbincludegraphics{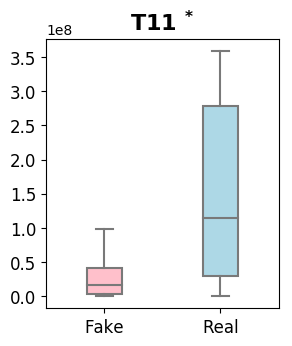}
\tcbincludegraphics[scale=1.05]{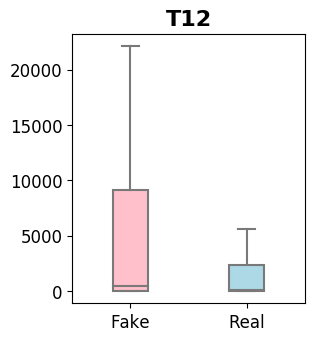}
\end{tcbraster}
\vskip -1em
\caption{The box plots demonstrating the differences in the distribution of temporal features of fake and real news pieces from PolitiFact dataset. Statistically significant features are represented by asterisk in the feature title. We observe similar patterns in Gossipcop dataset, and we omit the results due to the space limitation.
} 
\end{figure*}

We compare the temporal features of the macro-level propagation network of fake and real news in Figure \ref{fig:temp_politifact_boxplot} (from $T_1$ to $T_8$) and have the following observations:
\begin{itemize}
    \item The temporal features $T_{2}$, $T_{3}$, $T_{4}$, $T_{7}$ and $T_{8}$ from macro-level are statistically significant between fake and real news, under \textit{t}-test.
    
    \item The time difference between the first tweet and the last retweets ($T_{2}$) is smaller for fake news than real news. This indicates that fake news lives shorter than real news in social media on average in our datasets.
    
    \item Time difference between the first tweet and tweet with maximum outdegree ($T_{3}$) is smaller for fake news than real news in both datasets. It shows that fake news pieces are more likely to be shared earlier by an influential user than real news.
    
    \item Further, the time difference between the first and last tweet posting news ($T_{4}$) is shorter for fake news. This shows tweets related to fake news are posted in a shorter interval of time, which aligns with previous observations in~\cite{vosoughi2018spread} that tweets related to fake news are posted in a shorter interval of time and  spread faster than real news.
\end{itemize}

\subsection{Micro-Level Propagation Network}
Micro-level propagation networks involve users conversations towards news pieces on social media over time. It contains rich information of user opinions towards news pieces. Next, we introduce how to extract features from micro-level propagation networks from structural, temporal and linguistic perspectives.


\subsubsection{Structure analysis} : Structural analysis in the micro network involves identifying structural patterns in conversation threads of users who express their viewpoints on tweets posted related to news articles.

\begin{itemize}

\item ($\mathbf{S_{10}}$) \textit{Tree depth} : 
Depth of the micro propagation network captures how far is the conversation tree for the tweets/retweets spreading a news piece.


\item ($\mathbf{S_{11}}$) \textit{Number of nodes}: The number of nodes in the micro-level propagation network indicates the number of comments that are involved. It can measure how popular of the tweet in the root.



\item ($\mathbf{S_{12}}$) \textit{Maximum Outdegree}: In micro-network, the maximum outdegree indicates the maximum number of new comments in the chain starting from a particular reply node.

\item ($\mathbf{S_{13}}$) \textit{Number of cascade with with micro-level networks}: This feature indicates the number of cascades that have at least one reply.

\item ($\mathbf{S_{14}}$) \textit{Fraction of cascades with micro-level networks}: This feature indicates the fraction of the cascades that have at least one replies among all cascades.
\end{itemize}

The comparison of structural features for micro-level propagation networks of fake news and real news is demonstrated in Table~ \ref{tab:micro_struct}. We can see that:

\begin{itemize}
    \item Structural feature distributions of $S_{10}$, $S_{11}$, and $S_{14}$ are statistically different between fake news and real news in both datasets.
    \item The micro-level propagation networks of fake news is deeper ($S_{10}$) than real news significantly under t-test in both datasets, which is consistent with the observations in macro-level propagation networks revealed previously and in~\cite{vosoughi2018spread}.
    \item In addition, the fraction of cascades with micro-level networks ($S_{14}$) of fake news is greater than that of real news significantly under t-test in both datasets. The reason may be that fake news articles are more likely to be related to controversial and trending topics, which drives more engagements in terms of comments than real news articles.
\end{itemize}

\begin{table*}[!htbp]
\small
\label{tab:res}
\begin{center}
\caption{Statistics of linguistic features from micro propagation network. Stars denote statistically significant under $t$-test.}
\label{tab:micro_ling}
\begin{tabular}{| c | c | c | c| c| c| c | c |c | c|c|c|c|}
\hline
\textbf{Features } & \multicolumn{6}{ c |} {\textbf{PolitiFact}} & \multicolumn{6}{  c |}{\textbf{GossipCop}}\\ 
\cline{2-13}
& \multicolumn{3}{ c |}{\textbf{Fake}} & \multicolumn{3}{  c |}{\textbf{Real}} &  \multicolumn{3}{ c |}{\textbf{Fake}} & \multicolumn{3}{  c |}{\textbf{Real}}\\
\cline{2-13}
& \textbf{Min} & \textbf{Max} & \textbf{Avg}& \textbf{Min} & \textbf{Max} & \textbf{Avg} & \textbf{Min} & \textbf{Max} & \textbf{Avg}& \textbf{Min} & \textbf{Max} & \textbf{Avg}\\\hline
$L_{1}$ &0.0 & 6.0 & 1.12 & 0.0 & 15.0 & 1.305 & 0.0 & 23.5 & 1.094$^{*}$ & 0.0 & 20.0 & 1.010$^{*}$ \\
\hline
$\mathbf{L_{2}}$& -0.772 & 0.855 &0.007$^{*}$ & -0.585 & 0.894 & 0.045$^{*}$ &  -0.961 &  0.9607  &  0.051$^{*}$ &-0.922 & 0.969 & 0.077$^{*}$\\
\hline
$\mathbf{L_{3}}$ & -0.693& 0.855 &-0.001$^{*}$& -0.885  & 0.894&0.0428 $^{*}$ & -0.961 &  0.961 & 0.046$^{*}$ & -0.921 & 0.969 & 0.074$^{*}$ \\ 
\hline
$L_{4}$&-0.811 & 0.879 &  0.0156 & -0.934 & 0.664  & 0.027&  -0.960 & 0.938 & 0.044$^{*}$ & -0.893 & 0.969& 0.066$^{*}$ \\ 
\hline
$L_{5}$&  -0.811 & 0.879 &  0.011 & -0.934 &  0.851 & 0.028  & -0.961 &  0.993 & 0.039$^{*}$ & -0.896 &0.969 & 0.062$^{*}$ \\ 
\hline
\end{tabular}
\end{center}
\vskip -1em
\end{table*}
\subsubsection{Temporal analysis}
Micro-level propagation network depicts users' opinions and emotions through a chain of replies over time. The temporal features extracted from micro network can help understand exchange of opinions in terms of time. Following are some of the features extracted from the micro propagation network,

\begin{itemize}

\item ($\mathbf{T_{9}}$) \textit{Average time difference between adjacent replies in cascade}: It indicates how frequent users reply to one another.

\item ($\mathbf{T_{10}}$) \textit{Time difference between the first tweet posting news and first reply node}: It indicates how soon the first reply is posted in response to a tweet posting news.

\item ($\mathbf{T_{11}}$) \textit{Time difference between the first tweet posting news and last reply node in micro network}: It indicates how long a conversation tree lasts starting from the tweet/retweet posting a new piece.

\item ($\mathbf{T_{12}}$) \textit{Average time difference between replies in the deepest cascade}: It indicates how frequent users reply to one another in the deepest cascade.

\item ($\mathbf{T_{13}}$) \textit{Time difference between first tweet posting news and last reply node in the deepest cascade}: Indicates the life span of the conversation thread in the deepest cascade of the micro network.

\end{itemize}

The differences in the distribution of temporal features from micro-level networks of fake and real news are visualized in Figure \ref{fig:temp_politifact_boxplot} and we make the following observations:

\begin{itemize}
    \item The temporal features $T_{9}$, $T_{10}$ and $T_{11}$ for fake news and real news are statistically significant under \textit{t} - test for both datasets.
    
    \item The average time difference between adjacent replies $T_{9}$ is longer for fake news than real news, and it shows users take a longer time to respond to each other. The time difference between the tweet and the first reply  $T_{10}$ is shorter for fake news, which may indicate that users take less time to reply to tweets related fake and it takes more time to reply to another users comments.
    
    \item Further, the time difference between the first tweet posting news and the last reply node $T_{11}$ is shorter for fake news than real news and this clearly indicates users engage with fake news for a shorter interval of time and this is consistent with macro-level.
    
\end{itemize}

\subsubsection{Linguistic analysis}

People express their emotions or opinions towards fake news through social media posts, such as skeptical opinions, sensational reactions, etc. These textual information has been shown to be related to the content of original news pieces. Thus, it is necessary to extract linguistic-based features to help find potential
fake news via reactions from the general public as expressed
in comments from micro-level propagation network. 
Next, we demonstrate the sentiment features extracted from the comment posts, as the representative of linguistic features. We leave stance features as future work since existing tools on stance prediction generally exploit the similar set of features with sentiment~\cite{mohammad2017stance} and more nuanced aspect of opinions towards specific targets, which is not directly available in our datasets. 
We utilize the widely-used pre-trained model VADER~\cite{gilbert2014vader} to predict the sentiment score for each user reply, and extract a set of features related to sentiment as follows,

\begin{itemize}


\item ($\mathbf{L_{1}}$) \textit{Sentiment ratio}: We consider a ratio of the number of replies with a positive sentiment to the number of replies with negative sentiment as a feature for each news articles because it helps to understand whether fake news gets more number of positive or negative comments.

\item ($\mathbf{L_{2}}$) \textit{Average sentiment}: Average sentiment scores of the nodes in the micro propagation network. Sentiment ratio does not capture the relative difference in the scores of the sentiment and hence average sentiment is used.

\item ($\mathbf{L_{3}}$) \textit{Average sentiment of first level replies}: This indicates whether people post positive or negative comments on the immediate tweets posts sharing fake and real news.   
\item ($\mathbf{L_{4}}$) \textit{Average sentiment of replies in deepest cascade}:  Deepest cascade generally indicate the nodes that are most propagated cascade in the entire propagation network. The average sentiment of the replies in the deepest cascade capture the emotion of user comments in most influential information cascade. 
    
\item ($\mathbf{L_{5}}$) \textit{Sentiment of first level reply in the deepest cascade}: Deepest cascade generally indicate the nodes that are most propagated cascade in the entire propagation network. The sentiment of the first level reply indicates the user emotions to most influential information cascade.
\end{itemize}


 

\begin{table*} [htbp!]
\small
\vspace{-0.2cm}
\centering \caption{Best Performance comparison for fake news detection with different feature representations}
\begin{tabular}{|l|l|c|c|c|c|c|c|c|}
\hline
Datasets & Metric  &RST & LIWC &STFN&  HPNF & RST\_HPNF & LIWC\_HPNF&STNF\_HPNF   \\
\hline \hline
\multirow{4}{*}{\textbf{PolitiFact}} & Accuracy &0.796  &0.830 & 0.649 & 0.843 &0.875  &0.872  &  0.856\\
\cline{2-9}
&Precision &0.821 &  0.855 &   0.605 & 0.835  &  0.873& 0.869 &  0.809 \\
\cline{2-9}
&Recall &  0.752 & 0.792 & 0.836 & 0.851 & 0.876 &  0.872 & 0.927 \\
\cline{2-9}
&F1 & 0.785 &0.822  &0.702 &  0.843 & 0.875 &   0.871& 0.864\\
\hline
\hline
\multirow{4}{*}{\textbf{GossipCop}}   & Accuracy  & 0.600&  0.725 & 0.796 &  0.861  &0.861  & 0.869 & 0.863 \\
\cline{2-9}
&Precision & 0.603 & 0.773 &0.812 & 0.854 & 0.850 &0.856  &0.857 \\
\cline{2-9}
&Recall &0.586  & 0.637 &  0.770& 0.869 &0.876  &0.887  & 0.871  \\
\cline{2-9}
&F1 & 0.594 &  0.698 & 0.791 &  0.862  & 0.863  &0.871  &0.864  \\
\hline
\end{tabular} \label{tab:performance}
\end{table*}

We obtain the aforementioned linguistic features for micro-level propagation networks of fake news and real news in both Politifact and Gossipcop datasets. As shown in Table~\ref{tab:micro_ling}, we analyze the distribution of linguistic features and have the following observations: 

\begin{itemize}
    \item The linguistic features $L_2$, and  $L_3$ are significantly different for fake news and real news in both datasets.
    \item The average sentiment of replies ($L_2$) to  fake news is lower than the average sentiment of replies to real news in both the datasets under statistic t-test. It shows that tweets related to fake news receive more negative sentiment comments over real news. A similar result is observed in the sentiment of comments posted directly to tweets captured by feature $L_{3}$.
\end{itemize}


\section{Evaluating Propagation Features}\label{sec:eval}

In this section, we address \textbf{RQ2}. We explore whether the hierarchical propagation network features can help improve fake news detection, and how we can build effective models based on them. Moreover, we perform feature importance and model robustness analysis. We first introduce how to represent the hierarchical propagation network features $\mathbf{f}_i$ for a news item $a_i$. Let $\mathcal{G}_i$ denote the temporal propagation network of news piece $a_i$. For $\mathcal{G}_i$, we extract all types of propagation features and concatenate them into one feature vector $\mathbf{f}_i$. We also denote the proposed \textit{H}ierarchical \textit{P}ropagation  \textit{N}etwork \textit{F}eature vector $\mathbf{f}_i$ as HPNF.

\subsection{Experimental Settings}
To evaluate the performance of fake news detection algorithms, we use the following metrics, which are commonly used to evaluate classifiers in related areas: Accuracy (Acc), Precision (Prec), Recall (Rec), and F1. We randomly choose 80\% of news pieces for training and remaining 20\%
for testing, and the process is performed for 5 times and the average performance is reported. The details of baseline feature representations are given as below:

\begin{itemize}
\item \textbf{RST}~\cite{ji2014representation}: RST can capture the writing style of a document by extracting the rhetorical relations systematically. It  learns a transformation from a bag-of-words surface representation into a latent feature representation~\footnote{The code is available at: https://github.com/jiyfeng/DPLP}. 

\item \textbf{LIWC}~\cite{pennebaker2015development}: LIWC extracts lexicons that fall into different psycholinguistic categories, and learn a feature vector through multiple measures for each document~\footnote{The software and description of measures are available at: http://liwc.wpengine.com/}.

\item \textbf{STFN}~\cite{vosoughi2018spread}: STFN includes the structural and temporal features proposed in~\cite{vosoughi2018spread} for macro-level propagation network, i.e., tree height, number of nodes, max breadth of the tree, fraction of unique users, time taken to reach depth 1 in propagation, time taken to reach depth of 2 in propagation, number of unique users within level 1 and number of unique users within level 3 of propagation.


\item \textbf{RST\_HPNF}. RST\_HPNF represents the concatenated features of RST and HPNF, which includes features extracted from both news content and hierarchical propagation network.

\item \textbf{LIWC\_HPNF}. LIWC\_HPNF represents the concatenated features of LIWC and HPNF, which includes features extracted from both news content and hierarchical propagation network.

\item \textbf{STNF\_HPNF}. STNF\_HPNF represents the concatenated features of STNF and HPNF, which includes features structural and temporal features discussed in STNF and hierarchical propagation network features.

\end{itemize}
Note that for a fair and comprehensive comparison, we choose
the above feature extraction methods from following aspects: (1) \textbf{news content}, such as RST and LIWC; and (2) \textbf{propagation network}, such as Structure and Temporal features for Fake News Detection (STFN)~\cite{vosoughi2018spread}. We also combine RST, LIWC and STNF feature with HPNF to further explore if HPNF provides complementary information. For a fair comparison, we use the classifier that performs best on each feature set and compare the effectiveness of these different feature representations.

\subsection{Fake News Detection Performance Comparison}
We test the baseline features on different learning algorithms and choose the one that achieves the best performance (see Table~\ref{tab:performance}). The algorithms include Gaussian Naive Bayes (GNB for
short), Decision Tree (DT), Logistic Regression (LR), and Random Forest (RF). We used the open-sourced scikit-learn~\cite{pedregosa2011scikit} machine learning framework in Python to implement all these algorithms. To ensure a fair comparison of the proposed features and baseline features, we ran all the algorithms using default parameter settings. The experimental results are shown in Table~\ref{tab:performance}. We have the following observations:
\begin{itemize}
\item For news content-based methods, we see that LIWC performs  better than RST. This indicates that the LIWC vocabulary can better capture the deceptiveness in news content, which reveals that fake news pieces are very different from real news in terms of word choice from psychometrics perspectives.

\item Our proposed HPNF can achieve the best performance in both datasets on most of metrics compared with all other baseline methods. This shows that the extracted features from macro-level and micro-level propagation networks can help improve fake news detection significantly.

\item For propagation network-based methods, we can see that HPNF performs better than STNF consistently in both datasets. This is because HPNF includes features from micro-level networks from structural, temporal and linguistic perspectives that are useful for fake news detection. STNF only encodes the features from macro-level propagation network.

\item In addition, we observe that by combining HPNF features with existing features can further improve the detection performances. For example, RST\_HPNF performs better than either RST or HPNF, which reveals that they are extracted from orthogonal information spaces, i.e., RST features are extracted from news content and HPNF features from hierarchical propagation network on social media, and have complementary information to help fake news detection. We have similar observations for other features, i.e., (LIWC\_HPNF \textgreater LIWC, LIWC\_HPNF\textgreater HPNF) and (STNF\_HPNF\textgreater HPNF, STNF\_HPNF\textgreater STNF).




\end{itemize}


\begin{figure}[!tbp]
\centering
\subfigure[PolitiFact dataset]{
 {\includegraphics[scale=0.145]{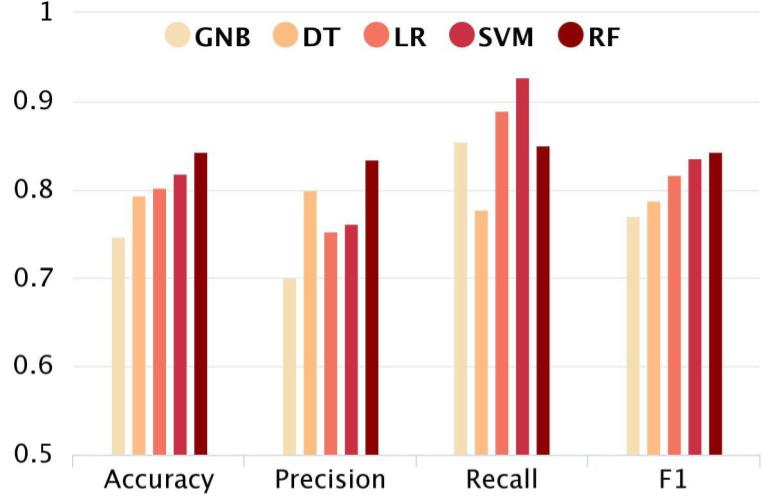}}
}
\subfigure[GossipCop dataset]{
 {\includegraphics[scale=0.16]{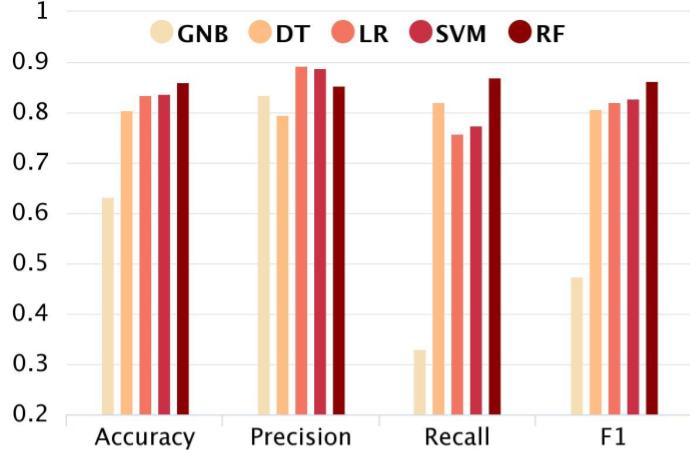}}
}
\vskip -1em
\caption{Detection Performance for HPNF with Different Learning Algorithms}
\label{fig:user_learning}
\vskip -1em
\end{figure}
We further evaluate the robustness of the extracted features HPNF. We demonstrate the fake news detection performances using different classifiers (see Figure~\ref{fig:user_learning}). These algorithms have different learning biases, and thus their performance is often  different for the same task. While we observe that: (1) RF achieves the best overall performance on both datasets; and (2) while the performance of RF is slightly better than other learning algorithms, the results are not significantly different across algorithms. This demonstrates that when sufficient information is available in the hierarchical propagation network features, the performance is not very sensitive to the choice of learning algorithms.

\subsection{Feature Importance Analysis}

In this subsection, we analyze the importance of the features in different granular levels to understand how each type of features contributes the prediction performance in fake news detection. We analyze
feature importance in the Random Forest (RF) by computing
a feature importance score based on the Gini impurity \footnote{https://bit.ly/2T1j29K}.

First, we evaluate the fake news detection performance on different levels of hierarchical propagation network including a) Micro-level; b) Macro-level; and c) both micro-level and macro-level (All) and compare their contributions to fake news detection in table \ref{tab:feature_group_level}. We have the following observations: (i) The combination of micro-level and macro-level features can achieve better performance than either micro-level or macro-level features in both datasets consistently. This shows that features from different levels  provide complementary information in feature dimension and thus help fake news detection; (ii) In general, we observe that micro-level features can achieve good performance, which demonstrates the necessitates of exploring micro-level features; (iii) Compared with macro-level features, micro-level features may not always perform better. For example, features from micro-level networks perform better than that from macro-level networks in PolitiFact dataset, and vice versa for GossipCop dataset.

\begin{table}[tbp!]
\centering
\caption{Best Detection Performance with Different Group of Features from HPNF} 
\begin{tabular}{|p{13.7mm}|p{13.7mm}|c|c|c|c|}
\hline
  Datasets & Level & Acc & Prec & Rec & F1\\
\hline
\multirow{3}{*}{PolitiFact} &Micro & 0.834 & 0.823 &0.847  &0.835\\
\cline{2-6}
&Macro & 0.807 &  0.816&  0.789 &0.802 \\

\cline{2-6}
& All &  \textbf{0.843} & \textbf{0.835} & \textbf{0.851} & \textbf{0.843} \\ 
\hline
\hline
\multirow{3}{*}{GossipCop} &Micro &  0.843 &0.841  & 0.845 &0.843 \\
\cline{2-6}
&Macro & 0.852 & 0.841 & 0.868 &  0.854\\
\cline{2-6}
 &All &  \textbf{0.861} & \textbf{0.854} & \textbf{0.869}  & \textbf{0.862}\\
\hline
\end{tabular} \label{tab:feature_group_level}
\vskip -1em
\end{table} 
\begin{table}[t]
\centering
\caption{Best Detection Performance with Different Group of Features from HPNF }
\begin{tabular}{|p{13.7mm}|p{13.7mm}|c|c|c|c|}
\hline
  Datasets & Type & Acc & Prec & Rec & F1\\
\hline
\multirow{4}{*}{PolitiFact} & Structural &0.681 & 0.681  & 0.672 & 0.676\\
\cline{2-6}
&Temporal &0.793 & 0.716 & \textbf{0.963} & 0.821 \\
\cline{2-6}
&Linguistic &0.659 & 0.648  &  0.683&0.665\\
\cline{2-6}
&All &  \textbf{0.843} & \textbf{0.835} & 0.851 & \textbf{0.843} \\
\hline
\hline
\multirow{4}{*}{GossipCop} & Structural &  0.826 &0.828  & 0.823 &0.826  \\
\cline{2-6}
&Temporal &  0.826& 0.827 & 0.825 & 0.826\\
\cline{2-6}
&Linguistic & 0.578  &  0.594& 0.491  &0.538 \\
\cline{2-6}
&All &  \textbf{0.861} & \textbf{0.854} & \textbf{0.869}  & \textbf{0.862}\\
\hline
\end{tabular} \label{tab:feature_group}
\vspace{-0.2cm}
\end{table}

Next, we evaluate the performance of different types of features from hierarchical propagation network including a) Structural; b) Temporal; c) Linguistic and d) combination of structural, temporal and linguistic (All), and compare their classification performance in Table \ref{tab:feature_group}. We have the following observations i) Temporal features perform better than both structural and linguistic features in both the datasets and this shows that temporal features have more importance in the classification task; ii) Structural features performs better than linguistic features in both the datasets as the micro network have limited linguistic contents; and iii) When the features from all perspectives are considered, the classification performance is better than considering either of the three features and this shows the features have complimentary information to differentiate fake news from real news.

\begin{figure}[h]
\centering{
\hspace{-0.2cm}
\includegraphics[scale=0.45]{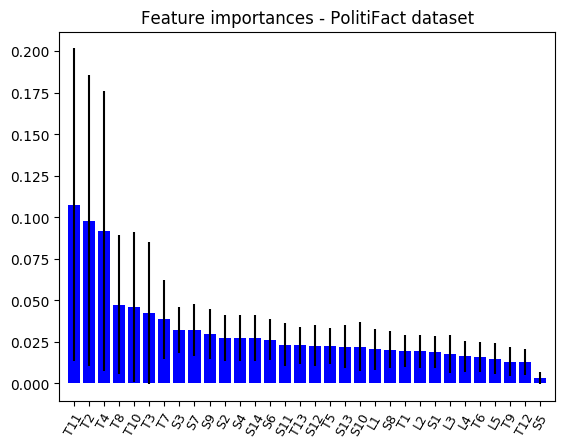}
\vskip -1em
\caption{Feature importance in Politifact dataset}
\label{fig:politifact_feat_importance}}
\end{figure}

\begin{figure}[t]
\vskip -1em
\centering{
\hspace{-0.2cm}
\includegraphics[scale=0.45]{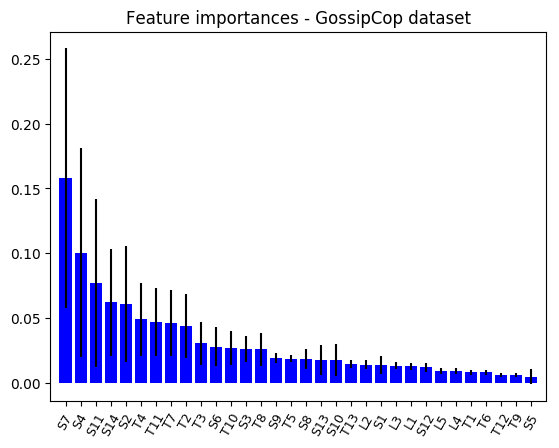}
\vskip -1em
\caption{Feature importance in GossipCop dataset}
\label{fig:gossipcop_feat_importance}}
\vskip -1em
\end{figure}

 From the Figure \ref{fig:politifact_feat_importance}, we can observe that: (i) the temporal features of PolitiFact dataset have higher importance scores over the structural and linguistic features; (ii) The feature $T_{11}$ shows that lifespan of the engagements in the micro network is the most important feature in fake news classification. Similarly, the life span of news in the macro network captured by $T_{2}$ shows the second highest importance score. This indicates that the longevity of fake and real news in social media is different; and (iii) Among structural features extracted from Politifact, the maximum out-degree of the macro network $S_{3}$ has more importance than other structural features. Figure \ref{fig:gossipcop_feat_importance} demonstrates the feature importance results on Gossipcop dataset. We make the following observations: (i) the fraction of cascades with retweets $S_{7}$ in the macro network has the highest importance score.  This shows difference in the scale of  spreading scope of fake and real news; (ii) In addition, the number of cascades in macro network $S_{4}$ has the second highest importance score; and (iii) the time difference between the first and last tweet posting news $T_{4}$ has higher importance score among temporal features. This confirms our findings that fake news tends to spread in a short period of time on social media than real news, and this actually serves an important feature for fake news detection.


\section{Related Work}\label{sec:related}
In this section, we introduce the related from two-folds: fake news detection and fake news propagation.
\subsection{Fake News Detection}
Fake news detection approaches generally fall into two categories: (1) using \textit{news content}; and (2) using \textit{social contexts}~\cite{shu2017fake}. For news content based approaches, features are extracted as linguistic-based such as writing styles~\cite{potthast2017stylometric}, and visual-based such as fake images~\cite{gupta2013faking}. 
Linguistic-based features capture specific writing styles and sensational headlines that commonly occur in fake news content~\cite{potthast2017stylometric}, such as lexical and syntactic features. Visual-based features try to identify fake images~\cite{gupta2013faking} that are intentionally created or capturing specific characteristics for images in fake news.
 News content based models include i) knowledge-based: using external sources to fact-checking claims in news content~\cite{magdy2010web,wu2014toward}, and 2) style-based: capturing the manipulators in writing style, such as deception~\cite{rubin2015truth} and non-objectivity~\cite{potthast2017stylometric}. 
 
Different from content-based approaches, social context based approaches incorporate features from social media user profiles, post contents, and social networks. User features can measure users' characteristics and credibilities~\cite{castillo2011information}. Post features represent users' social responses, such as stances~\cite{jin2016news}. Network features are extracted by constructing specific social networks, such as diffusion networks~\cite{kwon2013prominent} or co-occurrence networks~\cite{ruchansky2017csi}. All of these social context models can basically be grouped as either stance-based or propagation-based. Stance-based models utilize users' opinions towards the news to infer news veracity~\cite{jin2016news}. Propagation-based models apply propagation methods to model unique patterns of information spread~\cite{jin2016news}.

Existing approaches that exploit user social engagements simply extract features to train classifiers without a deep understanding of these features, which makes it a black-box that is difficult to interpret. Thus, we perform, to our best knowledge, the first in-depth investigation of various aspects of hierarchical propagation network for their usefulness for fake news detection.

\subsection{Fake News Propagation}
Diffusion-based models typically focus on modeling how fake news spreads/diffuses in the social network. The most notable work in this area is that of \cite{vosoughi2018spread} who analyze the diffusion of falsehoods (rumors) and truth on Twitter. Based on a large scale analysis \cite{vosoughi2018spread} claim that falsehoods tend to diffuse faster than truthful claims on social networks. Consequently, several works follow up on this line of work to comprehensively characterize the nature of fake news diffusion and dissemination~\cite{shao2017spread,babcockfake,bovet2018influence,wang2018cure}. ~\cite{shao2017spread} analyze the role of social bots in the diffusion of low credibility content and suggest that such automated accounts are particularly active in disseminating low-credibility content before the content becomes viral. Similarly ~\cite{babcockfake} note that not all fake news is the same and its effect on campaigns and communities differ. Consequently, they explore the reactions of different communities to fake news conversations on Twitter. ~\cite{wang2018cure} study the role of anonymity in the dissemination of online fake news. Intriguingly, their findings suggest that user identification and verification has limited utility in preventing the dissemination of fake news. 

Existing fake news propagation work mainly focuses on analyzing the macro-level propagation and does on perform an in-depth study on utilizing various propagation network features for fake news detection. To fill this gap, we construct a hierarchical propagation network from both macro-level and micro-level and exploit the features from structural, temporal and linguistic perspectives for fake news detection.

\section{Conclusion and Future Work}\label{sec:conclude}
In this paper, we aim to answer questions regarding the correlation between hierarchical propagation networks and fake news and provide a solution to utilize features from different perspectives from hierarchical propagation networks for fake news detection. Now we, summarize our findings of each research question and discuss the future work.

\textbf{RQ1} \emph{What are the characteristics of the structure, temporal and linguistic of hierarchical propagation networks of fake and real news?} To perform this study, we first construct the hierarchical propagation networks from macro-level and micro-level. For each type of network, we extract various features from structural, temporal and linguistic perspectives for fake news and real news. We compare these features to see if they are different or not for fake and real news with statistical analysis.

\textbf{RQ2} \emph{Can we use the extracted features to detect fake news and how?} With the quantitative analysis of news hierarchical propagation network features, we build different learning algorithms to detect fake news. We evaluate the effectiveness of the extracted features by comparing with several existing baselines. The experiments show that: (1) these features can make significant contributions to help detect fake news; (2) these features are overall robust to different learning algorithms; and (3) temporal features are more discriminative than linguistic and structural features and macro and micro level features are complimentary.

This work opens up the doors for many areas of research.
First, we can learn to predict whether a user will spread a fake news piece or not by studying the structures of hierarchical propagation networks, which is a precursor for mitigating fake news dissemination. Second, we can exploit the hierarchical structure of propagation networks to perform unsupervised fake news detection. Third, we may combine the extracted explicit propagation network features with deep learning models to further boost the performance of fake news detection.
\bibliographystyle{aaai}
\bibliography{refs}

\clearpage

\end{document}